\documentclass[pre,superscriptaddress,showkeys,showpacs,twocolumn]{revtex4-1}
\pdfoutput=1
\usepackage{graphicx}
\usepackage{epstopdf}
\usepackage{latexsym}
\usepackage{amsmath}
\begin {document}
\title {Entropic long-range ordering in an adsorption-desorption model}
\author{Adam Lipowski}
\affiliation{Faculty of Physics, Adam Mickiewicz University, Pozna\'{n}, Poland}
\author{Dorota Lipowska}
\affiliation{Faculty of Modern Languages and Literature, Adam Mickiewicz University, Pozna\'{n}, Poland}
\begin {abstract}  We examine a two-dimensional nonequilibrium lattice model where particles adsorb at empty sites and desorb when the number of neighbouring particles is greater than a given threshold. In a certain range of parameters the model exhibits entropic ordering similar to some hard-core systems. However, contrary to hard-core systems, upon incresing the density of particles the ordering is destroyed. In the heterogenous version of our model, a regime with slow dynamics appears, that might indicate formation of some kind of glassy structures.
\end{abstract}

\maketitle

\section{Introduction}
Models with hard-core interactions serve as an idealization of a number of important physical systems.  Indeed, various aspects of liquids \cite{hoover}, glasses \cite{berthier}, liquid crystals \cite{frenkel}, or certain adsorbates \cite{yakovkin} were successfully examined using hard-core models mainly by means of numerical simulations. Studies of particular importance are those related to the emergence of long-range ordering such as, for example, freezing of hard spheres \cite{alder} or of hard disks \cite{jaster}, which proceeeds via an intermediate hexatic phase.  Let us emphasize that hard-core interactions render the temperature irrelevant and ordering in such systems is of purely entropic origin \cite{dickman} with coverage (or pressure) as a control parameter.
In the computationally less demanding lattice hard-core systems, some more detailed insight into the ordering process is available. For example, on a square lattice, and when hard-core exclusion prevents nearest neighbors  of a given particle from being occupied, the ordering transition turns out to belong to the Ising model universality class \cite{kamieniarz,liu}, which is related to the double degeneracy of the ordered phase. When nearest- and next-nearest-neighbor repulsions are present, a four-fold degenerate columnar order is formed. Although it is more difficult to establish the nature of the ordering transition in this case \cite{levin},  most works suggest the Ashkin-Teller universality class \cite{deepak}.

Having in mind formation of some adsorbate structures such as, for example, He on graphite \cite{carlos} or on graphyne \cite{ahn}, or H on W \cite{doyen,petrova}, or O on Pt \cite{jensen}, we should take into account that equilibrium hard-core models provide only a very approximate description of these complex physical phenomena \cite{garcia}. An important process, which often accompanies adsorption and affects, for example, a surface diffusion \cite{privman} or various surface chemical reactions, is desorption \cite{vlachos,ertl}.  In certain statistical mechanics studies, the role of desorption in some equilibrium as well as nonequilibrium hard-core systems has already been examined \cite{liu}. In the present paper, we describe a nonequilibrium model where the desorption rather than the hard-core exclusion plays the primary role in the formation of an entropy stablized long-range order. What is, in our opinion, interesting is that the resulting ordered structures, and perhaps accompanying phase transitions, are the same as those  in the hard-core systems but the nature of the ordering process is much different: the ordered structures are destroyed when the density  of particles increases (not decreases, as in hard-core systems). Our work thus suggests that an alternative mechanism may play the role in the formation of entropy stabilized long-range ordering. 

\section{Model}
We examine a collection of particles, which adsorb at a two-dimensional surface, but  when a particle gets surrounded by too many neigbouring particles, it desorbes.
Thus, in a statistical mechanics fashion, in our model we have  $N$~particles distributed (without overlaps) over sites of a square lattice of linear size $L$ with periodic boundary conditions.  In an elementary step of the dynamics of our model, one selects randomly a particle and if it is unstable, it is relocated to one of the randomly selected empty sites. A particle is considered unstable if the number of particles on neighboring sites is greater than a given value~$k$. Some of the model characteristics are time dependent and we define the unit of time (1~MC step) as $N$~elementary steps (one step per particle). 

Let us emphasize that dynamics of our model shifts a constant number of particles ($N$) and desorption is always followed by adsorption which are thus not independent processes.  Such an approach bears some resemblance to the method of constant coverage ensemble used in the context of some surface-reaction models \cite{ziff}.
What is more important, is the lack of detailed balance in our model since a stable particle has a zero probability of desorption. Dynamics of our model might thus get trapped in an absorbing state where each particle is stable. In general, it is impossible to describe the stationary state of such models in terms of equilibrium Gibbs distributions and they belong to the realm of nonequibrium statistical physics. Models of this kind include some versions of the contact process \cite{tome,durrett,liu121,marro}, but might describe also some adsorption-desorption systems \cite{liu98,evans2000}.

The density of particles $\rho=N/L^2$ and the parameter~$k$ thus control the behavior of the model.  Of course, when the density~$\rho$ is sufficiently small, after a short transient each particle finds a stable position surrounded by at most $k$~neighbors. Such a state is an absorbing state of the dynamics.  Analogously, when $\rho$~is large, relocated particles are unlikely to find stable positions  that, in addition, do not destabilize their neighbors, and consequently, a fraction of particles are constantly  reshuffled.   
More interesting, and less obvious, is the behavior in an intermediate density range.  To examine it in more details, we carried out Monte Carlo simulations. We used two types of neighborhoods: (i) nearest neighbors (4 sites) and (ii) nearest and next-nearest neighbors (8 sites), and the results we obtained are presented in the next sections.

\section{Nearest-neighbor interactions}
\begin{figure}
\includegraphics[width=\columnwidth]{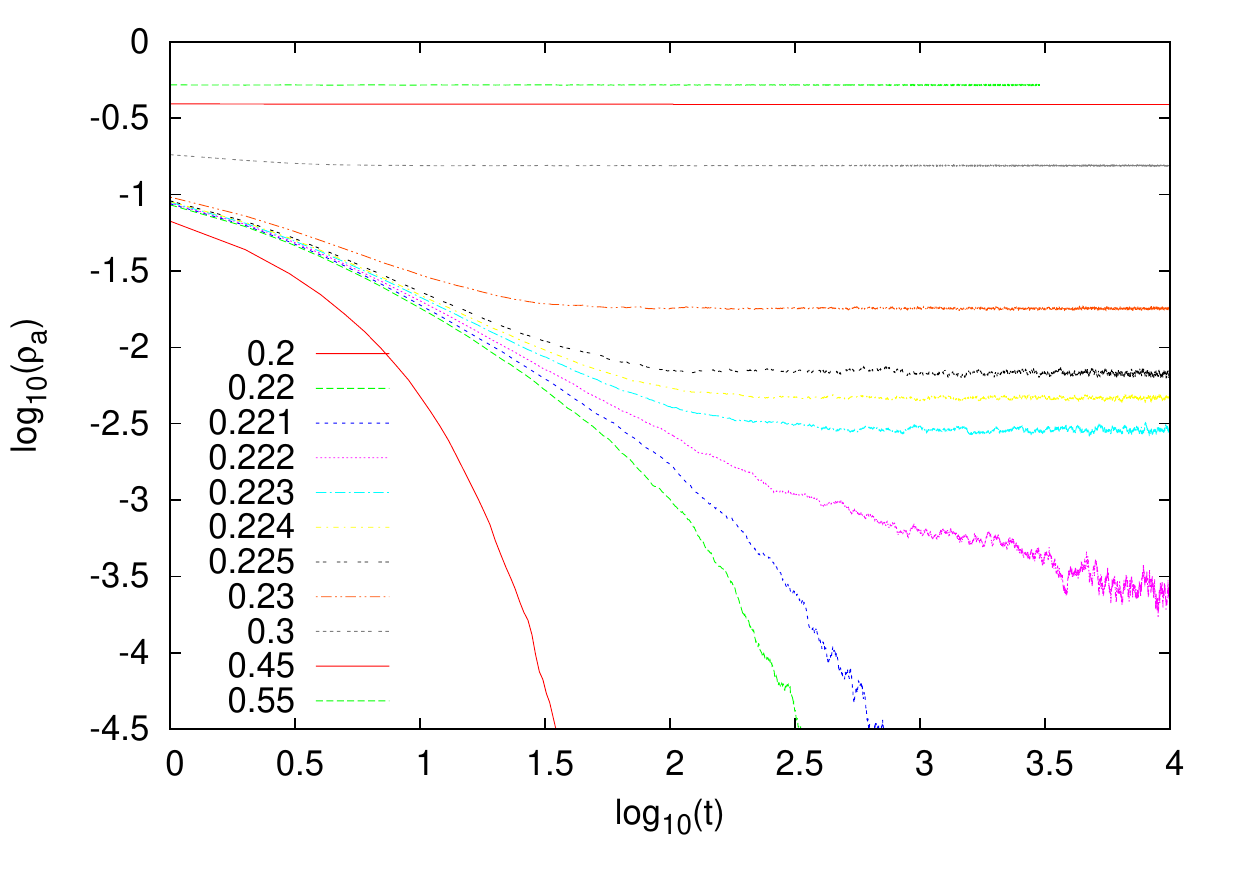}
\vspace{-0mm}
\caption{Time dependence of the density of active particles $\rho_a$ for the nearest-neighbor model with $k=0$ and $\rho=0.2$ (bottom curve), 0.22,..., 0.55 (top curve). Simulations were run  for $L=10^3$ and the results are averaged over 100 independent runs. The statistical error is of the order of noise seen in the plotted curves.}
\label{0times}
\end{figure}

\begin{figure}
\includegraphics[width=\columnwidth]{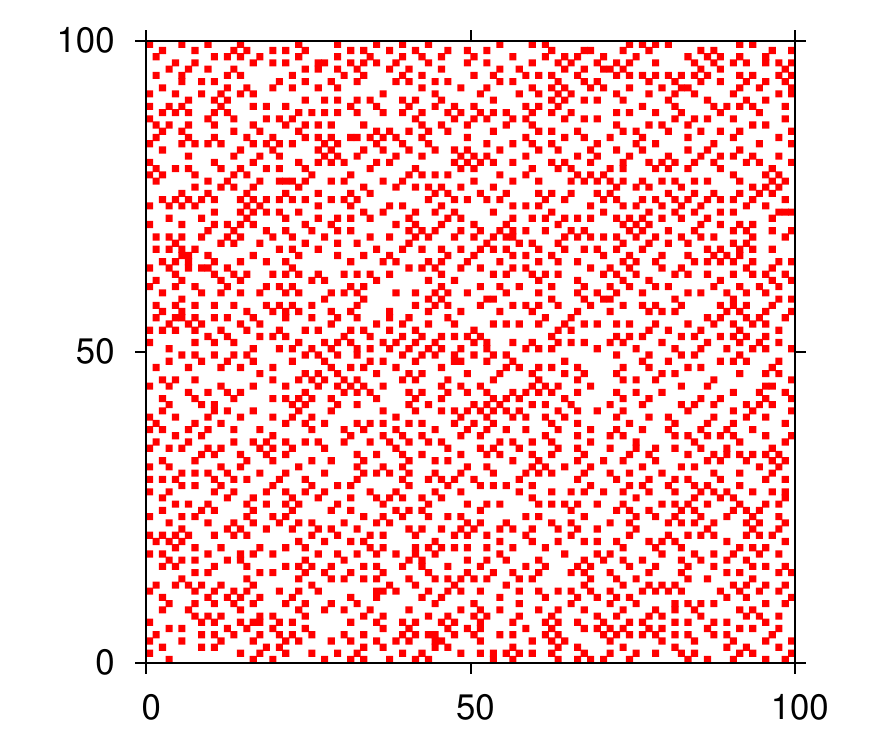}
\vspace{5mm}
\caption{The distribution of particles in the stationary state (after relaxation of a random initial configuration for $10^4$ MC steps) for the nearest-neighbor model with $k=0$ and $\rho=0.225$.}
\label{config225}
\vspace{10mm}
\end{figure}

\begin{figure}
\includegraphics[width=8cm]{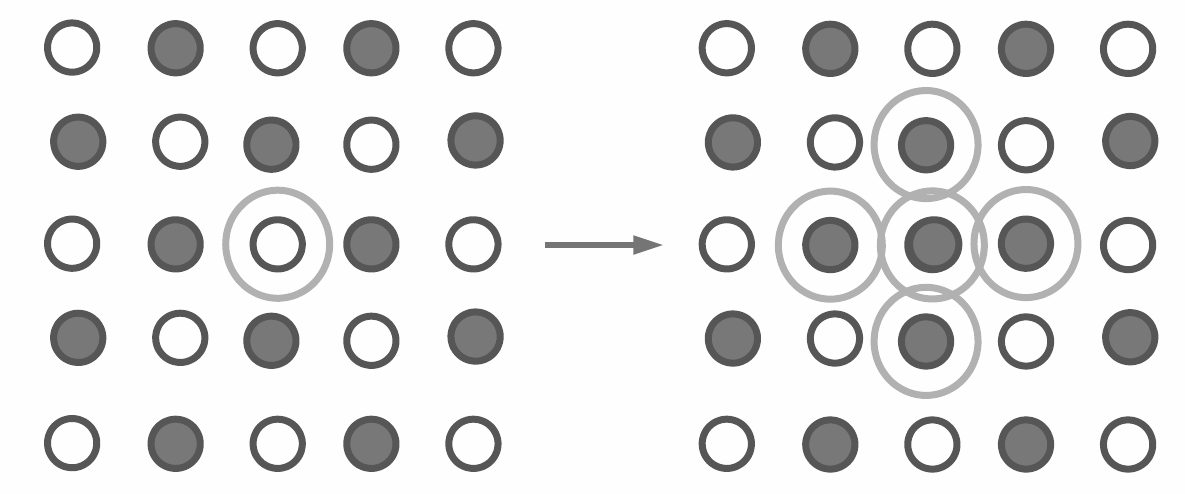}
\vspace{-0mm}
\caption{In the nearest-neighbor model with $k=0$, when an empty site in a cluster with a checkerboard structure gets occupied, it makes unstable all its four neighbors. A subsequent move is likely to erode the cluster ordering.}
\label{config0}
\end{figure}

In this case each site has four neighbors, which implies that $0\leq k \leq 4$. To introduce the methodology, we first examine $k=0$. After generating a random initial configuration with $N=\rho L^2$ particles, we redistribute them using the model dynamics. We calculated the average density of active (i.e., unstable) particles~$\rho_a$ as a function of time~$t$, and the results are presented in Fig.~\ref{0times}.
As expected, for small~$\rho$ ($\rho\leq 0.221$), we observe a fast decay of~$\rho_a$ and eventually the model reaches an absorbing state of the dynamics, where nearest neighbors of each particle are empty. On the other hand, for $\rho\geq 0.223$, the model remains in a state with a finite fraction of active particles. Let us notice that various periodic structures would satisfy the condition $k=0$ and the densest of them (with $\rho=0.5$) is the checkerboard ordering, analogous to the hard-core model with the nearest-neighbor exclusion \cite{kamieniarz,liu}. We do not observe formation of any of such global periodic structures and their dynamical creation is apparently unlikely. However, ordering might appear on a small scale. For example, for $\rho=0.225$ (Fig.~\ref{config225}), the snapshot configuration shows various clusters with a checkerboard pattern. Such clusters, however, are very unstable. If one of the empty sites is chosen and gets occupied, both this site as well as its four neighbors  turn into unstable sites (Fig.~\ref{config0}). As a result, large clusters of this kind do not form.

\begin{figure}
\includegraphics[width=\columnwidth]{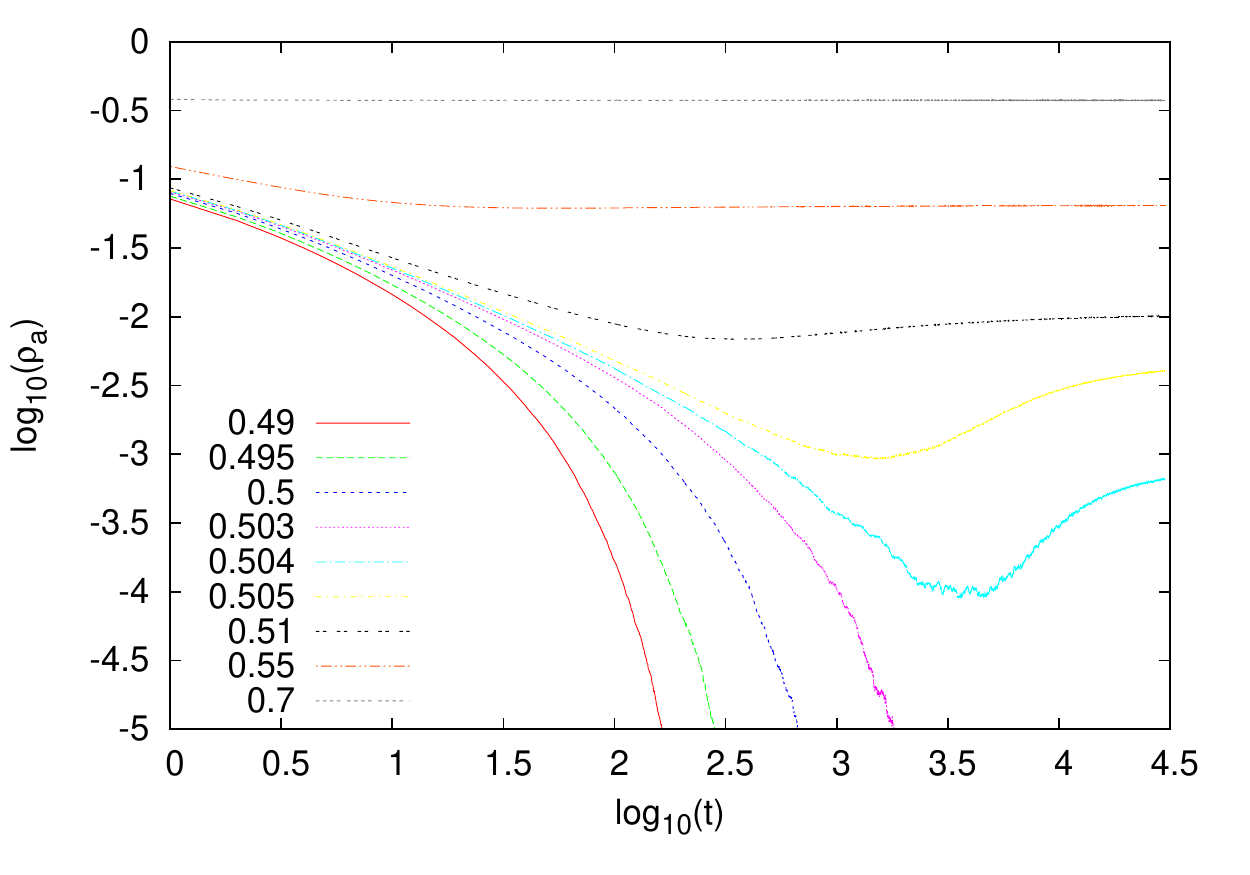}
\vspace{-0mm}
\caption{Time dependence of the density of active particles $\rho_a$ for the nearest-neighbor model with $k=2$ and $\rho=0.49$ (bottom curve), 0.495,..., 0.7 (top curve). Simulations were run  for $L=10^3$ and the results are averaged over 100 independent runs.}
\label{2time}
\end{figure}

\begin{figure}
\includegraphics[width=\columnwidth]{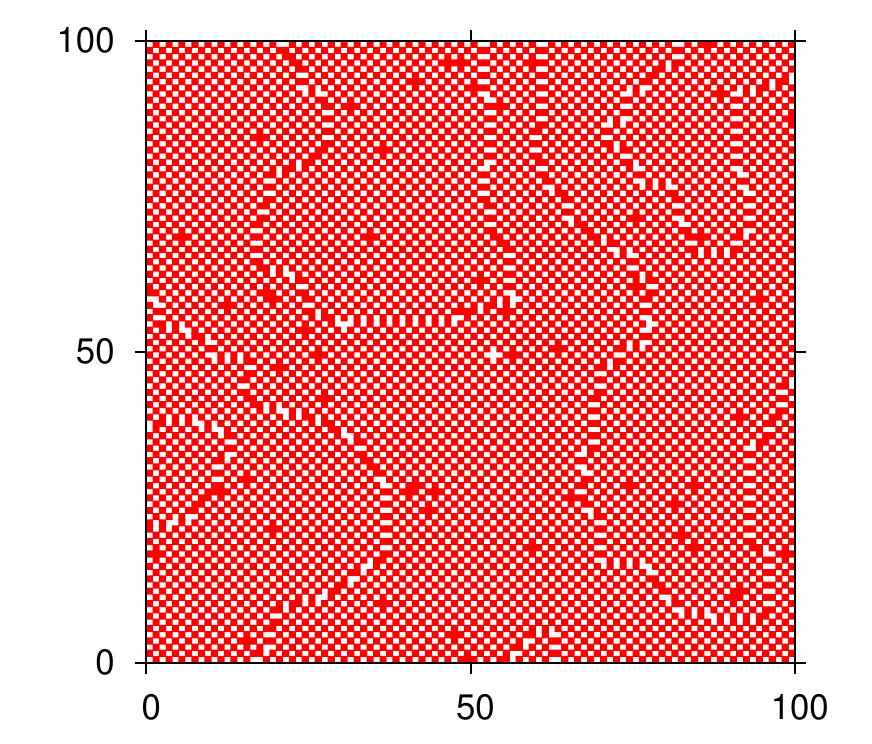}
\vspace{-0mm}
\caption{The distribution of particles in the stationary state (after relaxation of a random initial configuration for $10^4$ MC steps) for the nearest-neighbor model with $k=2$ and $\rho=0.51$.}
\label{2config51}
\vspace{10mm}
\end{figure}

\begin{figure}
\includegraphics[width=8cm]{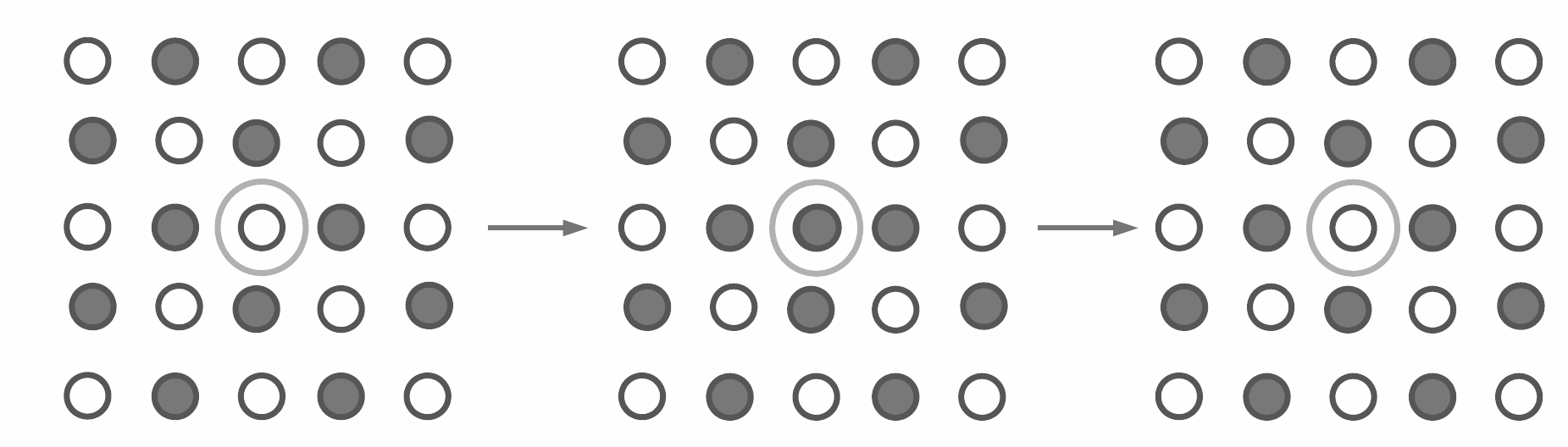}
\vspace{5mm}
\caption{In the nearest-neighbor model with $k=2$, the empty site in the checkerboard structure that gets occupied is the only unstable site. The inital  configuration is likely to be restored unless some other nearby empty site gets occupied.}
\label{config1}
\end{figure}

\begin{figure}
\includegraphics[width=4cm]{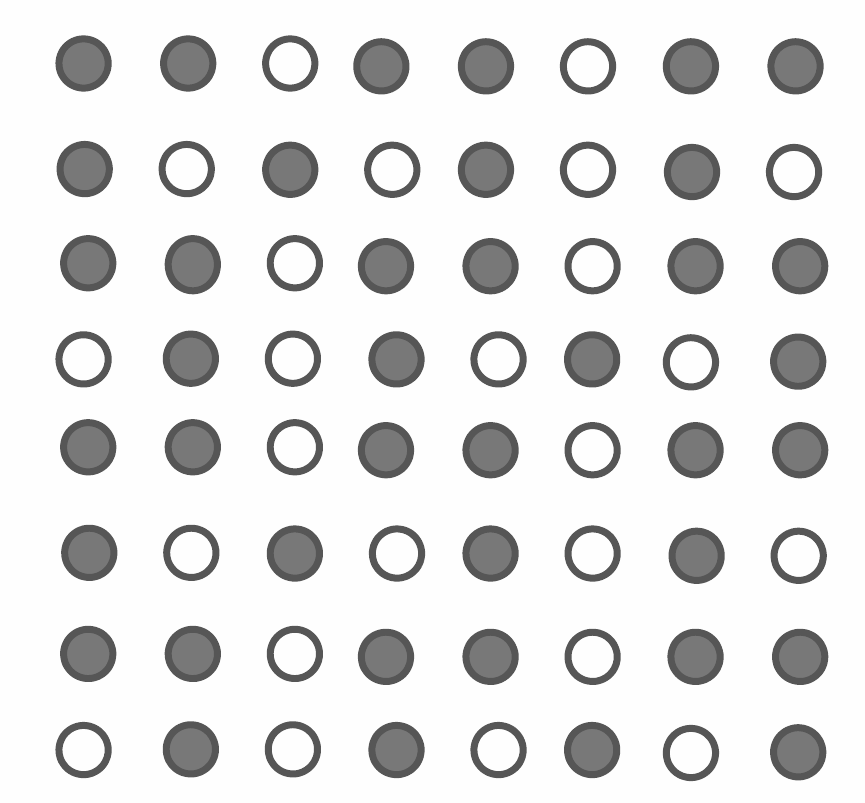}
\vspace{-0mm}
\caption{Periodic structure with the density $\rho=7/12$, where particles satisfy the limit $k=2$ in the nearest-neighbor version, and $k=4$  in the next-nearest-neighbor version.}
\label{config}
\end{figure}

For $k=2$, a much different scenario takes place. In this case, the transition between absorbing and active regimes of the model takes place around $\rho=0.5035(10)$ (Fig.~\ref{2time}). While the absorbing state, similarly to the case $k=0$,  is disordered, the active regime is different. Indeed, for $\rho=0.51$ formation of long-range ordered structures is clearly seen in Fig.~\ref{2config51}. Let us notice that a checkerboard structure satisfies actually the stronger limit $k=0$, since in this case the number of occupied nearest neighbors is zero. It implies a stronger stability of such structures: an empty site that gets occupied does not destabilize its neighbors (Fig.~\ref{config1}). One can easily find higher-density periodic structures that satisfy the limit $k=2$ (Fig.~\ref{config}), but they are not dynamically stable (as is the checkerboard structure in the $k=0$ case) and  we did not observe their formation during the evolution of our model.

\begin{figure}
\includegraphics[width=\columnwidth]{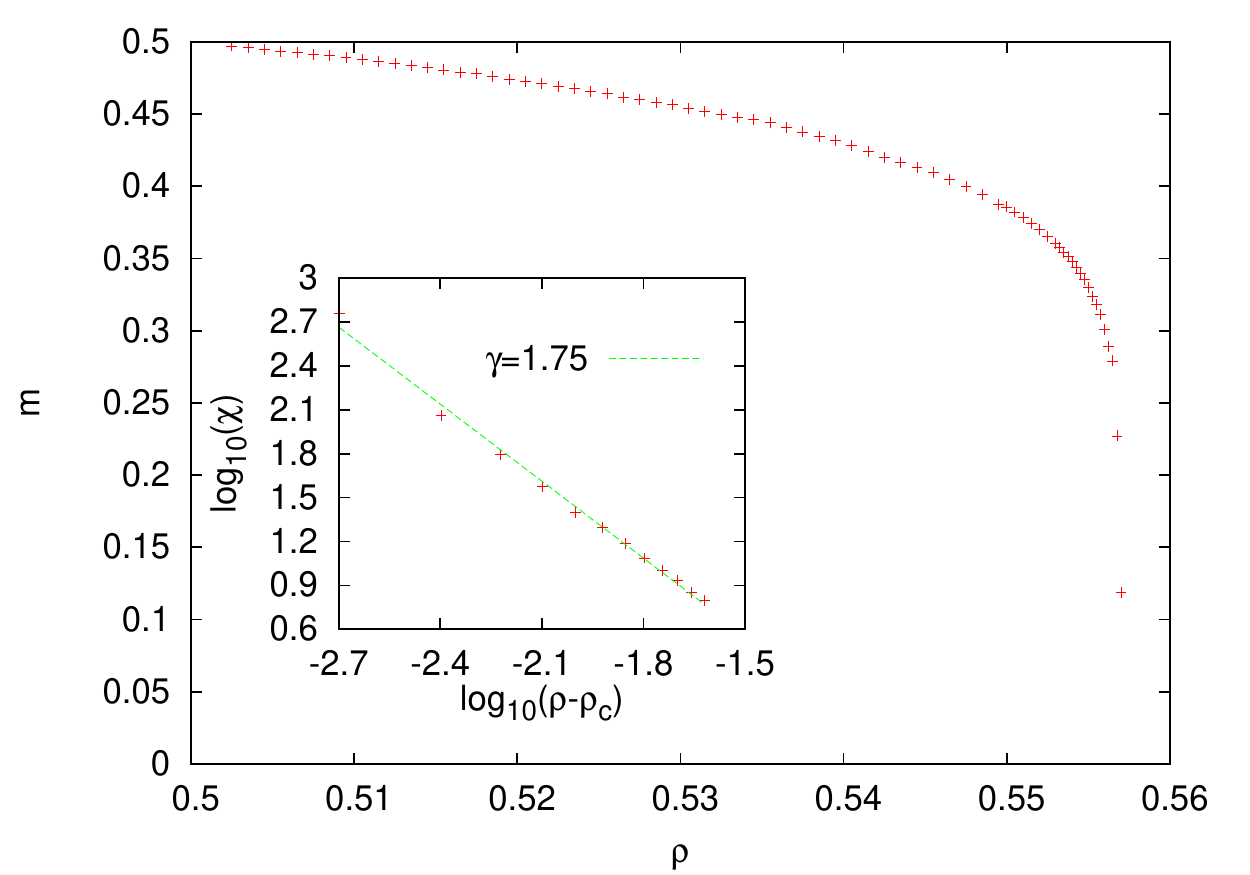}
\vspace{-0mm}
\caption{The order parameter $m$ as a function of the density $\rho$ for the nearest-neighbor model with $k=2$ ($L=10^3$).  The inset suggests that the variance of the order parameter $\chi_m$ diverges at the critical point $\rho=0.556$ with the exponent $\gamma=1.75$.}
\label{2order}
\end{figure}

Upon increasing the density~$\rho$, the number of unstable particles~$\rho_a$ in the steady state also increases, wich gradually destroys the long-range ordering. To examine the process in more detail, we carried out simulations for $\rho>0.5$, which started from the predefined checkerboard ordering. We divided the lattice into two sublattices, A~and~B, and placed $L^2/2$ particles on the sublattice~A, while the remaining ones were randomly distributed on the sublattice~B. Running the model dynamics, we relaxed the system until it reached the steady state and then we measured the order parameter~$m$ defined as
\begin{equation}
m=\frac{1}{L^2}\left( \sum_{i\in A} n_i-\sum_{i\in B} n_i \right),
\label{m}
\end{equation}
where $n_i=0{\rm\ or\ }1$ for a site~$i$ being empty or occupied by a particle, respectively. The results  (Fig.~\ref{2order}) show that $m$ decays to~0 at $\rho=\rho_c\approx 0.556$. Assuming the power-law decay ($m\sim (\rho_c-\rho)^{\beta}$), we estimate $\beta\approx 0.17(5)$, and the fit is based on data close to the critical point ($0.555<\rho<0.556$). Taking into account that the checkerboard structure is double-degenerate,  one might expect that the transition at $\rho_c$ belongs to the Ising model universality class and the obtained estimate of $\beta$ is marginally consistent with the Ising model value $0.125$. We also measured the variance $\chi_m$ of the order parameter~$m$ in the $\rho>\rho_c$ regime (simulations started from a random initial configuration). The results (inset in Fig.~\ref{2order}) show that $\chi$ has a power-law divergence $\chi_m\sim (\rho-\rho_c)^{-\gamma}$ with $\gamma\approx 1.75$, which is in a very good agreement with the Ising model value. Presented numerical results are obtained for the system size $L=10^3$. Except the very vicinity of the critical point ($\rho=\rho_c$) the examined systems seem to be sufficiently large and the finite size effects are negligable. More precise estimations of the critical behaviour would certainly require more systematic analysis of finite size effects.

We carried out some simulations for $k=1$ and $k=3$, and we did not observe formation of a long-range ordering. The behavior of the model in these cases seems to be similar to the $k=0$ case. As a final remark in this section, let us notice that the decay of ordering via the Ising-like phase transition takes place upon the density increase. This is opposite to the behavior of hard-core systems, where the high-density phase is long-range ordered.

\section{Next-nearest-neighbor interactions}
\begin{figure}
\includegraphics[width=\columnwidth]{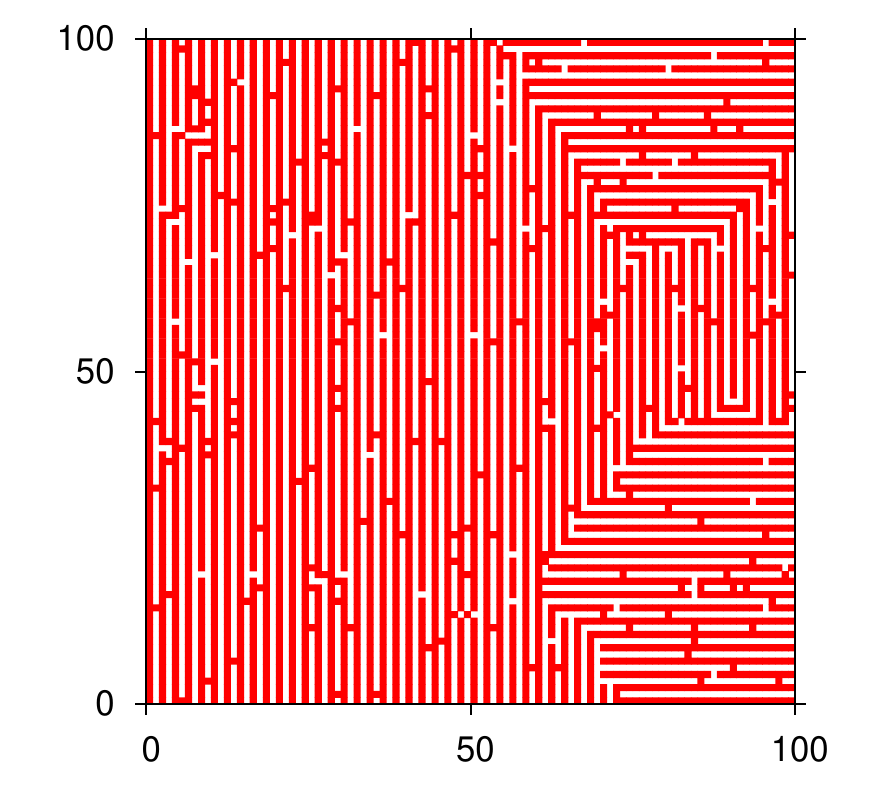}
\vspace{-0mm}
\caption{The distribution of particles in the stationary state (after relaxation of a random initial configuration for $10^4$ MC steps) for the nearest- and next-nearest-neighbor model with $k=4$ and $\rho=0.51$.}
\label{4config51}
\end{figure}

\begin{figure}
\includegraphics[width=\columnwidth]{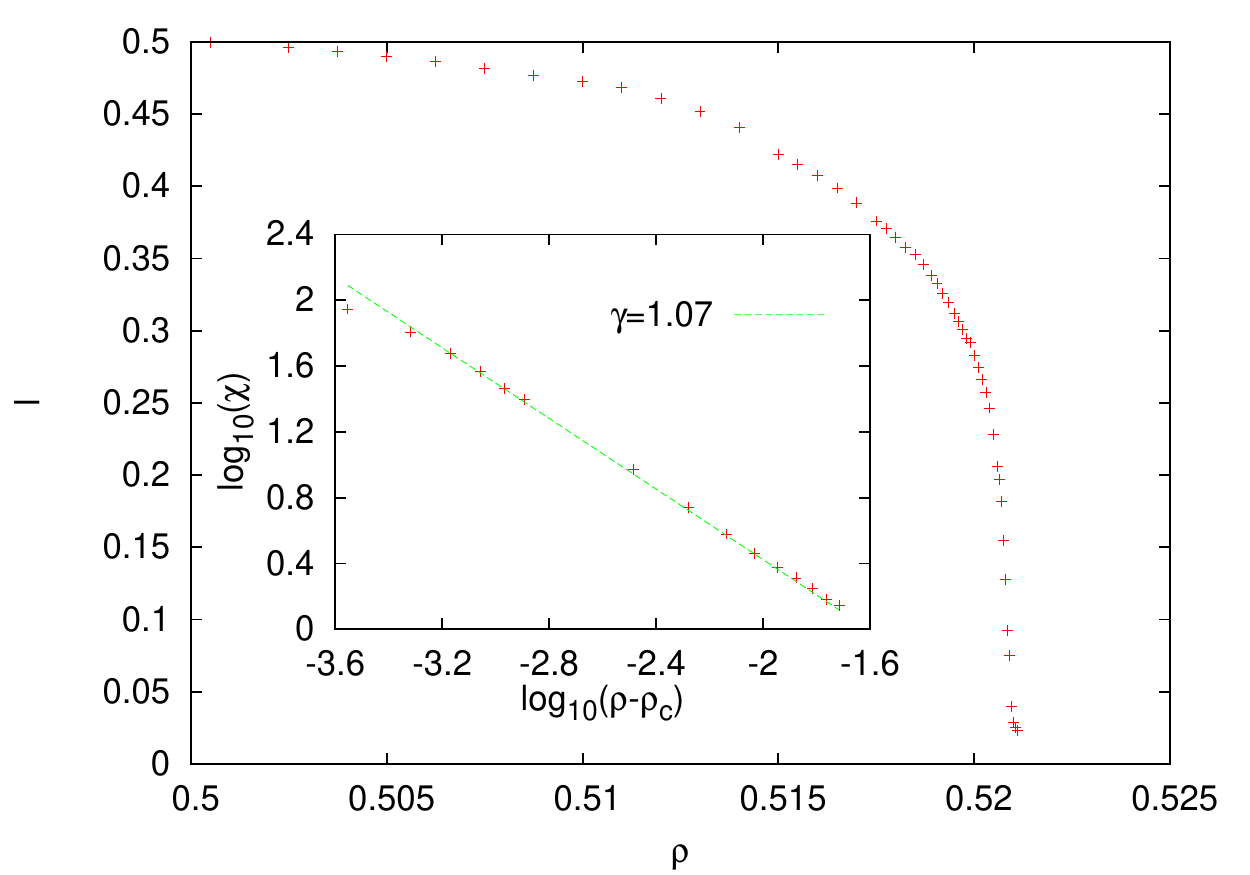}
\vspace{-0mm}
\caption{The columnar order parameter~$l$ as a function of the density $\rho$ for the next-nearest-neighbor model with $k=4$ ($L=10^3$).  The inset shows that the variance of the order parameter $\chi_l$ diverges at the critical point $\rho=0.5207$ with the exponent $\gamma=1.07$.}
\label{4order}
\end{figure}

We also carried out  simulations for the model with the nearest- and next-nearest-neighbor interactions, where each site has 8 such neighbors. The formation of  long-range ordering was observed only for $k=4$ and densities greater, but not much, than 0.5. The snapshot configuration for $\rho=0.51$ (Fig.~\ref{4config51}) clearly shows the formation of the columnar ordering. To examine such structures in more detail, we ran simulations with initial configurations with a predefined columnar ordering, similarly to the nearest-neighbor $k=2$ version. We measured the  columnar order parameter~$l$, which basically counts the number of sites with horizontally versus vertically placed neighbors:
\begin{equation}
l=\frac{1}{L^2} \sum_{i} l_i,
\label{l}
\end{equation}
where $l_i=1$ (or $-1$) for the site $i$, which has its two horizontal (or vertical) neighbors occupied  (otherwise $l_i=0$). Our numerical results show (Fig.~\ref{4order}) that similarly to the nearest-neigbour version, $l$ takes the maximum value 1/2 (perfect columnar ordering) at $\rho=0.5$. When  $\rho$ increases,  the number of unstable particles also increases, which gradually destroys an ordering.
The least-square fitting to the numerical data close to the transition point gives $\rho_c=0.5207(5)$ and $\beta=0.25(5)$, and the fit was made using data for $0.52<\rho<0.5207$. Moreover, from the behavior of the variance $\chi_l$ of the order parameter (inset in Fig.~\ref{4order}), we estimate $\gamma=1.07(3)$. The four-fold degeneracy of the columnar ordering suggests that, similarly to some hard-core systems with a columnar ordering \cite{deepak,blote2011}, the critical behavior of our model may belong to the Ashkin-Teller universality class. In such a case, one expects $\beta=1/12$  and $\gamma=7/6$ \cite{creswick}, and our estimate of $\gamma$ is very close to the expected value. The deviation of $\beta$ might be related to strong finite-size effects or the fact that the true asymptotic regime was not yet reached in our simulations. More detailed analysis would be clearly desirable.

\section{Nearest-neighbor interactions with heterogeneities}
Heterogeneity of size, mass, or shape of particles is known to play an important role in hard-core systems. For example, it might lead to the phase separation of different particles \cite{dijkstra} or to the formation of multiple glassy phases \cite{voitman,zacarelli}. Studying analogous phenomena in lattice models, which are usually computationally more tractable, might provide a valuable insight into the role of space dimension, range of interactions or symmetries. Despite decades of intensive research, the formation of a glassy state is  a particularly challenging problem. While its existence is well documented in the three-dimensional systems \cite{pusey,berthier}, the status of a two-dimensional glass is not certain. Although some works report certain dynamical glassy features in two-dimensional systems \cite{doliva}, some other question the  existence of a glassy transition in such systems \cite{donev,santen}. It is not our objective to address these important general questions but rather to show that a heterogeneous version of our model, which  may mimick the bi- or polydisperse hard-core systems, develops some slowly evolving characteristics which could suggest some relations with glassy systems.

\begin{figure}
\includegraphics[width=\columnwidth]{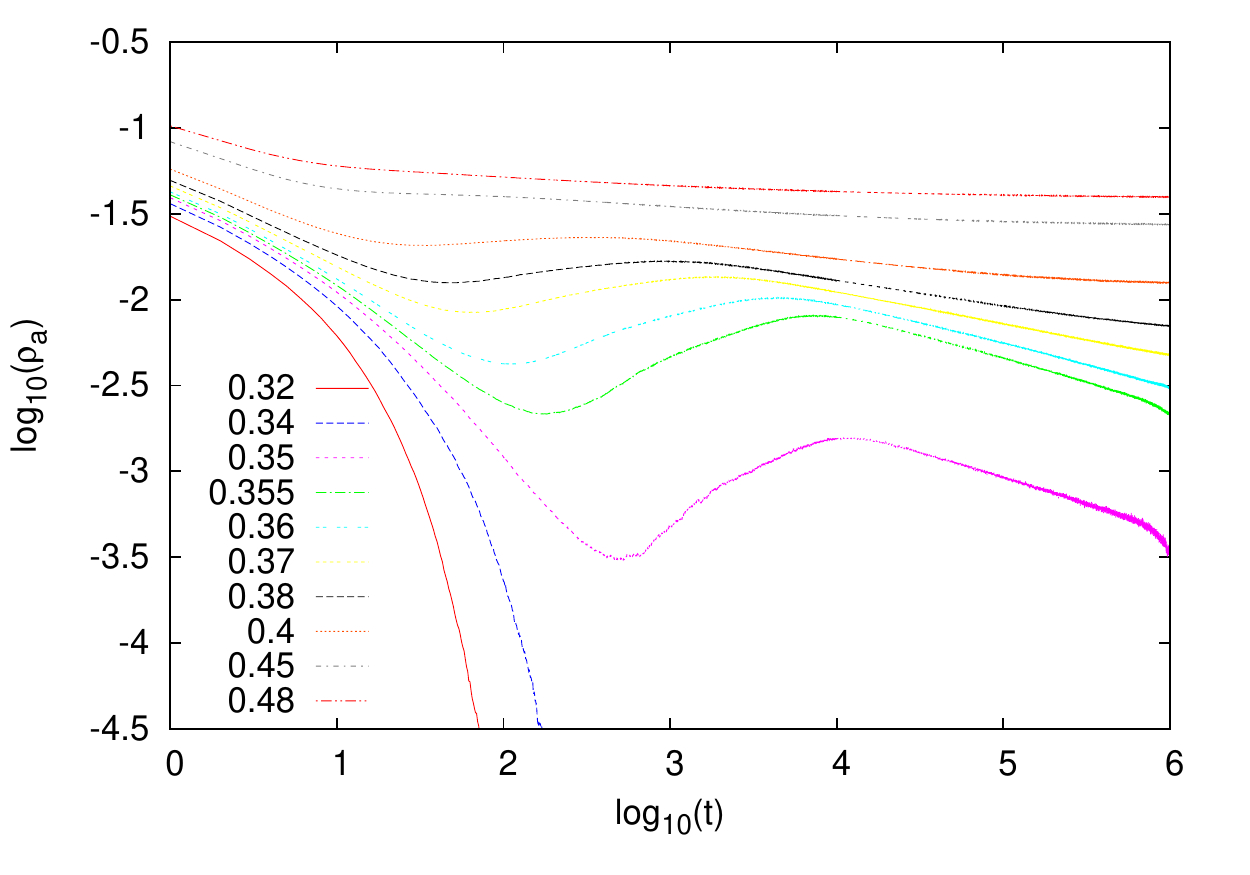}
\vspace{-5mm}
\caption{Time dependence of the density of active particles $\rho_a$ for the heterogeneous nearest-neighbor model  with $k=2 (90\%)$, and $k=0 (10\%)$ and for $\rho=0.32$ (bottom curve), 0.34,..., 0.48 (top curve). Simulations were run  for $L=300$ and the results are averaged over 100 independent runs.}
\label{mixture-active}
\end{figure}

\begin{figure}
\includegraphics[width=\columnwidth]{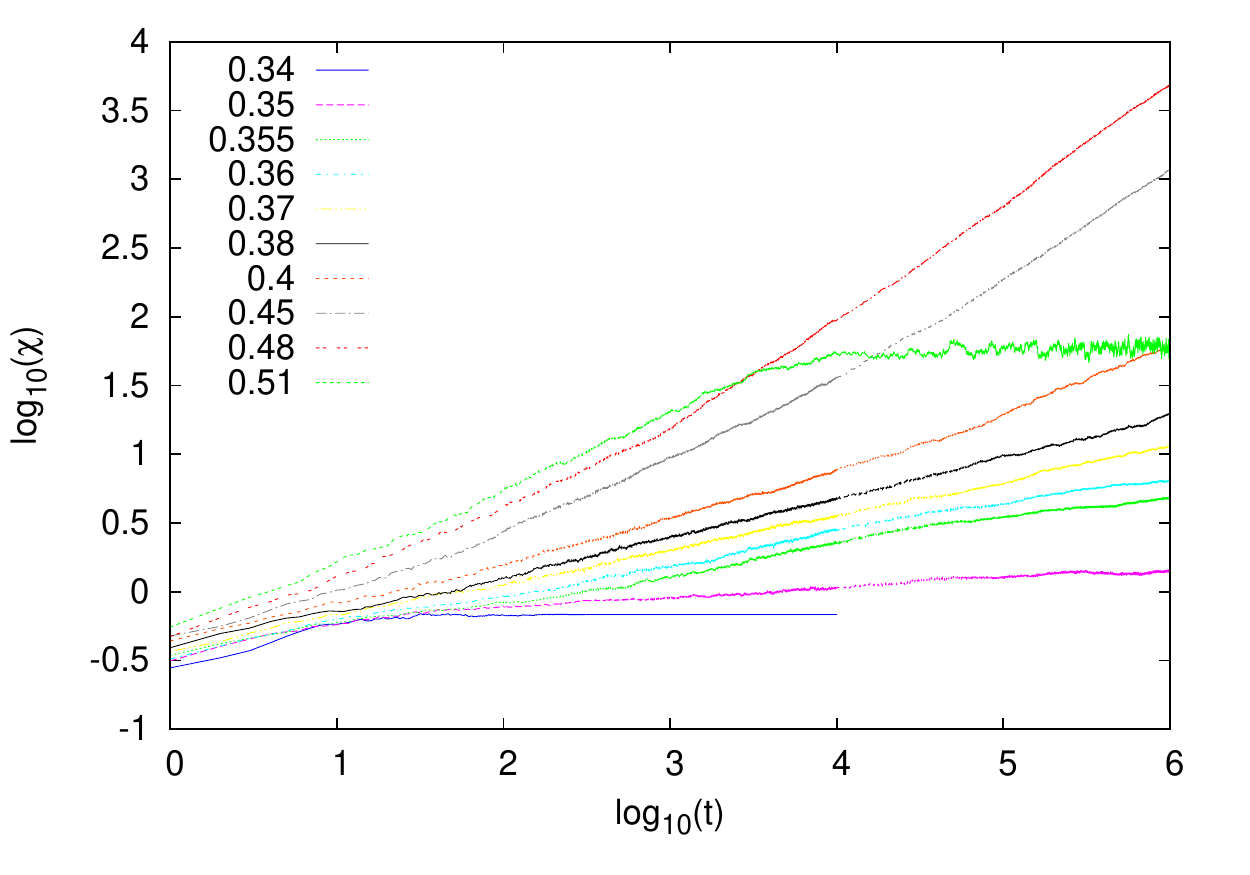}
\vspace{-0mm}
\caption{Time dependence of the variance of the order parameter $\chi_{m}(t)$ for the heterogeneous nearest-neighbor model  with $k=2 (90\%)$, and $k=0 (10\%)$ and for $\rho=0.34$ (bottom curve), 0.35,..., 0.51 (top curve, that bends horizontally around $t\sim 10^3$). Simulations were run  for $L=300$ and the results are averaged over 100 independent runs.}.
\label{mixture-susc}
\end{figure}

In particular, we examine a heterogeneous version of our nearest-neighbor model, where a fraction $p$ of particles obeys the dynamical rule with $k=2$, and the remaining fraction ($1-p$) with $k=0$. Simulations for $p=0.9$ show that $k=0$ particles hinder reaching the absorbing state and an active regime extends up to $\rho\sim 0.4$ (Fig.~\ref{mixture-active}).  Moreover, the absorbing regime seems to be separated into two sub-regimes. For lower densities ($\rho=0.32, \ 0.34$),  an ordinary, fast  (presumably exponential) decay of the density of active sites $\rho_a$ can be seen. However, for larger densities ($\rho=0.35\sim 0.37$), a  much slower decay of $\rho_a$ can be clearly seen. For example, for $\rho=0.355$ from the estimation of the asymptotic slope of the numerical data, we obtain $\rho_a\sim t^{-0.25}$. To examine in more detail the structure of the model, we calculated the time-dependent variance $\chi_m$ of the order parameter (\ref{m}) (Fig.~\ref{mixture-susc}).   For a moderately large density of particles ($\rho=0.45\sim 0.48$), the variance $\chi_m$ rapidly increases in time, which indicates formation of long-range ordered checkerboard structures. For larger density ($\rho=0.51$), the density of active particles $\rho_a$ is too large, which destroys a long-range order and $\chi_m$ saturates at a finite value. In the homogeneous case ($p=1$), we did not calculate the time dependent $\chi_m$ but an analogous behavior would be observed. Also similarly to the homogeneous case, at a low density of particles ($\rho=0.34$), the variance $\chi_m$ saturates at a small value, which indicates an absence of long-range ordered structures. Less evident is the behavior for intermediate densities ($\rho=0.35\sim 0.37$), where a noticeable but slow increase of $\chi_m$ can be seen. It may indicate a very slow growth of domains, thus providing a further evidence that in this regime the model exhibits some glassy characteristics.

\section{Conclusions}
In the present paper, we introduced a  simple adsorption-desorption model that may generate an entropic long-range ordering. The structures formed (checkerboard, columnar) bear some similarity to the entropic order in hard-core systems,  but the mechanism that generates ordering in our model is much different. In particular, the ordered phase exists for a certain intermediate particle density and gets destroyed upon a density increase---not upon decrease as in ordinary hard-core systems. The order-disorder transitions are likely to belong to the expected universality classes. For the double degenerate checkerboard ordering, the transition belongs to the Ising model universality class, and for the columnar ordering, our data are marginally consistent with the Ashkin-Teller universality class.  Precise estimation of critical exponents in our model would certainly require more careful analysis of statistical errors and of finite size effects. The main objective of the present paper is, however,  to present a new mechanism of entropic ordering in lattice models and its further analysis is postponed for the future.

Simulations show  that in the nearest-neighbor version, where each site has four neighbors, the checkerboard ordering appears for $k=2$. In the next-nearest-neighbor version with 8 neighbors, we found the columnar ordering  for $k=4$. One of the questions is why an ordering appears only for $k$ equal to the half of the number of neighbors. It might be related to the fact that in both cases the ordered phase originates at (or very close to)  $\rho=0.5$, but let us notice that for both types of the ordering, even the stronger stability criterion is satisfied and particles have less than $k$ occupied neighbors. One might thus imagine that, in principle, the dynamics with  smaller $k$ could reproduce such an ordering as well, but apparently this is not the case. Most likely, for smaller $k$, ordered structures are not dynamical attractors of the model, however, more convincing evidence of such a scenario would be desirable. Having in mind some adsorbing systems, it would be certainly more realistic to move an unstable particle with a diffusive dynamics rather than place it on a randomly selected empty site. Preliminary calculations (results will be presented elsewhere) show, however, that a similar long-range ordering should form also   in such a version. The present dynamics, where a desorbed particle hopes to randomly chosen empty site, might be more suitable in the context of some evaporation-recondensation systems. In such a case, however, temperature dependent effects should be taken into account.

Another issue, which in our opinion is worth further studies, is a slow dynamics in a heterogeneous version of our model with particles having different values of $k$. Simulations show that in such a case a new regime appears, where the evolution toward the absorbing state is very slow (but power-law). One might hope that a mixture of particles with different values of $k$ or with different ranges of interactions will lead to even slower dynamics, which would be more relevant in the context of glassy systems. A glassy state often appears in various hard-core systems and a heterogeneity (e.g., polydispersity) is known to enhance it. The dynamics of our model in some cases exhibits a considerable slowdown and its further studies may contribute to a better understanding of a somewhat unclear status of glassy dynamics in two-dimensional systems.


\end {document}